\documentclass[journal,comsoc]{IEEEtran}
\usepackage[T1]{fontenc}
\usepackage[dvipsnames,svgnames,x11names]{xcolor}

\usepackage[utf8]{inputenc}
\usepackage{acronym}
\usepackage{graphicx}
\usepackage{float}
\usepackage{tikz}
\usepackage{framed}
\usepackage{comment}
\usepackage{gensymb}
\usepackage{multirow}
\usepackage[hyphens]{url}

\usepackage{graphicx} 

\usepackage[english]{babel}

\newcommand{\filippo}[1]{}
\newcommand{\micropmu}{$\mu$-\ac{PMU}~}
\newcommand{\micropmus}{$\mu$-\acp{PMU}~}
\newcommand{\synchrophasor}{$\bm{\xi}(t_i)$}

\acrodef{ADC}{Analog to Digital Converter}
\acrodef{C-IoT}{Cellular-Internet of Things}
\acrodef{DER}{Distributed Energy Resource}
\acrodef{DFT}{Discrete Fourier Transform}
\acrodef{eDRX}{enhanced Discontinued Reception}
\acrodef{KPI}{Key Performance Indicator}
\acrodef{GPSDO}{GPS Digital Oscillator}
\acrodef{mMTC}{massive Machine Type Communication}
\acrodef{IoT}{Internet of Things}
\acrodef{ISR}{Interrupt Service Routine}
\acrodef{m-MTC}{massive Machine Type Communication}
\acrodef{NBIoT}{Narrow Band \ac{IoT}}
\acrodef{PDC}{Phasor Data Concentrator}
\acrodef{PMU}{Phasor Measurement Unit}
\acrodef{PV}{Photo-Voltaic}
\acrodef{PWM}{Pulse-Width Modulation}
\acrodef{ROCOF}{Rate of Change of Frequency}
\acrodef{TVE}{Total Vector Error}
\acrodef{UE}{User Equipment}

\usepackage{cite}

\ifCLASSINFOpdf
\else
\fi
\usepackage{amsmath}
\usepackage[cmintegrals]{newtxmath}
\usepackage{bm}

\usepackage{datetime}

\ifCLASSOPTIONcompsoc
  \usepackage[caption=false,font=normalsize,labelfont=sf,textfont=sf,subrefformat=parens,labelformat=parens]{subfig}
\else
  \usepackage[caption=true,font=footnotesize,subrefformat=parens,labelformat=parens]{subfig}
\fi
\usepackage{url}

\newcommand\copyrighttext{%
  \footnotesize \textcopyright 2021. This work has been submitted to the IEEE for possible publication. Copyright may be transferred without notice, after which this version may no longer be accessible.
  }
\newcommand\copyrightnotice{%
\begin{tikzpicture}[remember picture,overlay]
\node[anchor=south,yshift=4pt] at (current page.south) {\fbox{\parbox{\dimexpr\textwidth-\fboxsep-\fboxrule\relax}{\copyrighttext}}};
\end{tikzpicture}%
}

\begin{document}

\title{Experimental End-To-End Delay Analysis of LTE cat-M With High-Rate Synchrophasor Communications}

\author{Sureel~Shah,~Sayan~Koley,~and~Filippo~Malandra,~\IEEEmembership{Member,~IEEE,}
\thanks{S. Shah, S. Koley, and F. Malandra are with the Department
of Electrical Engineering, State University of New York at Buffalo, Buffalo, NY, 20156, USA. Corresponding author: Filippo Malandra, filippom@buffalo.edu.}}
\markboth{Preprint submitted to an IEEE Journal, 2022}%
{Shah \MakeLowercase{\textit{et al.}}: Bare Demo of IEEEtran.cls for IEEE Communications Society Journals}

\maketitle

\begin{abstract}
Micro-Phasor Measurement Units ($\mu$-PMUs) are devices that permit monitoring voltage and current in the distribution grid with high accuracy, thus enabling a wide range of smart grid applications, such as state estimation, protection and control. These devices need to transmit the synchronous measurements of voltage and current, also known as synchrophasors, to the power utility control center at high rate. The use of wireless networks, such as LTE, to transmit synchrophasor data is becoming increasingly popular. However, synchrophasors are included in small frames and it would be more efficient to use low power cellular solutions, such as LTE cat-M. In this work, we present experimental  research  on the  deployment  of  a  $\mu$-PMU with the ability to connect over a commercial LTE cat-M network. The deployed $\mu$-PMU is built with off-the-shelf hardware, such as Arduino microcontrollers, and is used to transmit data---compliant with the IEEE C37.118.2 standard---at a variable rate from 1 frame/s to 80 frames/s. A detailed network performance analysis is carried out to show the suitability of LTE cat-M to support $\mu$-PMU communications. Experimental results on performance indicators, such as delay and jitter, are reported. The effect of the LTE cat-M access mechanism on the time distribution of frame arrivals is also thoroughly analyzed.
\end{abstract}


\IEEEpeerreviewmaketitle

\copyrightnotice

\section{Introduction}\label{section:introduction}

\IEEEPARstart{P}{hasor} Measurement Units (\acsu{PMU}s) have traditionally been used to monitor the power grid and represent one of the major drivers in the transition from the old centralized power grid into the modern, diverse, and distributed smart grid.
These devices permit the computation of the so-called \textit{synchrophasors}, which are \textit{time-synchronized} estimations of the magnitude and phase of voltage and/or current waveforms. \acp{PMU} also measure other quantities, such as frequency, and \ac{ROCOF}, and are able to provide estimates of active and reactive power \cite{IEEE_27.244}. \ac{PMU} data are extremely time-sensitive and their transmission is mainly regulated by two sets of standard procedures, namely IEC 61850 \cite{brunner08}  and IEEE C37.118.2 \cite{IEEE_C37.118.2}. 
Synchrophasor data are employed in a number of smart grid applications, which mainly differ in the reporting rate and in the communication requirements, such as delay and frame error rate. A thorough overview of the most common \ac{PMU} applications along with their reporting and communication requirements is provided in \cite{vonmeier17}, where applications are categorized based on the required time resolution and latency. 
Latency spans across a wide range of values: for example, protection and control applications require a maximum latency in the order of a few ms, whereas other applications, such as state estimation or outage management, can tolerate network delays in the order of 1 s. Traditional \acp{PMU}, mainly used in the transmission grid \cite{duan20}, provided a \acf{TVE} precision of $\pm 0.01\%$, a phase angle accuracy of $\pm 0.003\degree$, a magnitude resolution of $\pm 0.0002 \%$ and an angle resolution of $\pm 0.001 \degree$ with around $120$ frames per second \cite{synchro}.




In recent years, a need for higher precision measurements in the distribution systems and micro grids entailed the deployment of a new type of synchrophasor measurement device, i.e., the micro-PMU (or $\mu$-PMU). This type of devices is characterized by a \ac{TVE} precision of $\pm 0.05\%$,  a phase angle accuracy of $\pm 0.001\degree$, a magnitude resolution of $\pm 0.0002 \%$ and an angle resolution of $\pm 0.002 \degree$ with up to $100-120$ frames per second \cite{psl17}. 
In the last decade, with the increased use of distributed generation elements, such as photovoltaic and wind farms, \micropmus have proliferated and need to be connected to the power grid in order to collect their synchrophasor data.

\acp{PMU} and \micropmu are connected to \ac{PDC}, which are in charge of collecting synchrophasors from several \acp{PMU}, aligning them, and then forwarding them to the power utility \cite{IEEE_27.244}. Traditionally, wired connectivity, such as Ethernet or power line communications (PLCs), have been preferred due to their high performance and low delay. However, in recent years, wireless networks are becoming an increasingly popular solution to connect \micropmus to the Internet thanks to their flexibility, ease of installation and scalability. Previous work on wireless networks for \acp{PMU} or \micropmus has predominantly focused on cellular technologies, such as WiMax, UMTS \cite{borghetti17} and LTE \cite{dervivskadic16,malandra18}. However, the coverage of WiMax is not extensively available, consequently undermining large-scale \micropmu deployments. On the other hand, UMTS and LTE, despite providing a quasi-ubiquitous coverage, were conceived to support human traffic, which is considerably different from \micropmu traffic \cite{samoilenko20}. One basic difference lies in the size of the synchrophasor frames, which are in the order of a few tens of Bytes, as opposed to the larger data commonly exchanged in human traffic such as video streaming or web surfing. As a result, the large resources required for LTE transmission, in terms of bandwidth and energy, would not be efficiently used to transmit \micropmu data. 

Therefore, in this work, we are proposing to wirelessly connect \micropmus using LTE cat-M, one of the so-called \ac{C-IoT} technologies \cite{accurso21}. This LTE-based technology is characterized by a limited use of the frequency spectrum and a lower cost, both in terms of hardware (LTE cat-M radio are cheaper than standard LTE radios) and in terms of data plans, which are cheaper with respect to other cellular networks. In order to test the suitability of LTE cat-M to transmit synchrophasor data, we deployed a \micropmu using Arduino microcontroller boards and a GPS module. This prototype of \micropmu is connected to an AC power signal generator, samples the power signal, generates synchrophasor  frames (compliant with the IEEE C37.118.2 standard), and transmits them to a laptop, which represents the \ac{PDC}, using a commercial LTE cat-M network in the US. To the best of our knowledge, this is the first study to propose using LTE cat-M for \micropmu communications. 

The main contributions of this paper are: i) the development of a low-cost low-complexity prototype of a $\mu$-PMU with LTE-M transmission capability, ii) the characterization of frames and transmission rates compliant to the IEEE C37.118.2 standard, and iii) an experimental analysis of end-to-end delay of LTE cat. M for synchrophasor communications.
The remainder of this paper is structured as follows: in Section \ref{section:stateoftheart}, an overview of the state of the art is presented; in Section \ref{section:system}, the proposed system modelling is described; numerical results are reported in Section \ref{section:results}; finally, the conclusions of this research are discussed in Section \ref{section:conclusions}.

\section{State of the Art}\label{section:stateoftheart}

The pervasive installation of \acp{DER} into power distribution systems along with the popularity of demand-response initiatives have led distribution networks to work outside their operational limits. To cope with the more stringent requirements on these systems, recent trends show that improvements of the distribution system management are preferred over network capacity increase \cite{uddin17}. \acp{PMU} and \micropmus are key elements to increase the observability and enhance metering in distribution networks \cite{aguero19}. These devices are expected to bring about substantial improvements in power systems in terms of reliability, security, and overall performance \cite{aguero19}.

There is a great deal of literature analyzing the positive effect of synchrophasor devices on the performance of distribution systems~\cite{mohsenian18,chai15,yang17,pignati15,gharavi15}.
In particular, Mohsenian et al provided examples to demonstrate the importance of using \acp{PMU} and \micropmus in the distribution grid \cite{mohsenian18}. They showed, in one of these examples, how to use data from multiple \acp{PMU} to identify a short-circuit fault event, therefore allowing a quick deployment of field crews at fault location. They also advocated the importance of developing predictive analytics based on machine learning and big data techniques~\cite{mohsenian18}. In~\cite{chai15}, Chai et al. proposed to use \acp{PMU} to enable real-time state estimation. To cope with the lack of a solid and steady communication infrastructure, the authors introduced C-DAX, a communication platform based on the information-centric networking theory. C-DAX employs a combination of PLC and optical fiber to support synchrophasor communications~\cite{chai15}. A real-time monitoring infrastructure was also proposed by Pignati et al. in~\cite{pignati15}, based on a hierarchical architecture with Ethernet, single-pair high-speed digital subscriber line and optical fiber. The proposed system was deployed on the EPFL campus and an average end-to-end latency of 65 ms was obtained. A similar hierarchical architecture was considered in 
\cite{gharavi15} with a combination of wired and wireless technology. Wireless links were emulated using the well-known EMANE platform\footnote{Networks and Communication Systems Branch: EMANE. Available: {http://cs.itd.nrl.navy.mil/work/emane/index.php}.}.

The use of wireless networks to support synchrophasor communications has become increasingly popular in recent years \cite{khan12, dervivskadic16,pourramezan17,pourramezan20,pourramezan20new,malandra18}. In \cite{khan12}, it was proposed to use WiMax technology for the wide area \ac{PMU} communications. The authors employed existing network simulation software (i.e., OPNET) in order to evaluate different scheduling services in a network with a maximum of 50 \acp{PMU}, a reporting rate of 25 Hz and a maximum application delay of 40 ms. Network simulation for \ac{PMU} communications was also performed in \cite{malandra18}, where the authors considered a smart-city environment with a realistic LTE cellular infrastructure. The employed network simulator\footnote{Available at \url{www.trafficm2modelling.com}.}, permitted considering up to 21 PMUs and 25 thousands \acp{PMU} along with traffic generated from more than 300 thousands smart meters; however, human traffic was not considered \cite{malandra18a,malandra17}. 
A more realistic study with LTE was carried out in \cite{dervivskadic16}, where Dervivskadic et al. used a commercial LTE network to support data transmission from 10 \acp{PMU}. The authors analyzed the total \ac{PDC} reporting latency with 4 different logics and then compared the experimental LTE results against a wired network benchmark with optical fiber. 

Even though latest generations of wireless networks provide promising results, the variability of wireless transmissions entails the possibility of dropping frames or unpredictable degradation of the performance, leading to large delays and potentially compromising the success of the considered power system applications. In order to cope with this effect, \cite{pourramezan17} proposed a statistical method to compensate for bad communication data. Pourramezan et al. later proposed a novel method to coordinate different \acp{PDC} to mitigate the variability of data transmitted by different \acp{PMU} over a commercial LTE network \cite{pourramezan20new}. Zhu et al. proposed in \cite{zhu20} to use spatial-temporal correlations of \acp{PMU} data to to fix missing measurements and abnormal values.

\begin{table*}[h!]
\caption{\label{tab:table1} Comparison of experimental PMU implementations.} 
\centering
\setlength{\tabcolsep}{7pt} 
\renewcommand{\arraystretch}{1.5} 
\resizebox{2\columnwidth}{!}{
\begin{tabular}{p{1.9cm}||ccccc}
PMU Name & GridTrak PMU \cite{gridtrak} & OpenPMU V1 \cite{openpmu} & OpenPMU V2 \cite{openpmu} & Mohapatra \cite{Mohapatra15} & Our Design   \\ \hline
\hline
Approx. cost & 150\$ & 400\$ & 100\$ & 500\$ & 150\$\\ \hline
Hardware & dsPIC30F microcontroller & NI DAQ & BeagleBone B. & Arduino Due & Arduino Mega\\ \hline
Software & C\# & LabView & Linux & Linux, Python & Python\\\hline
Open-Source & Yes & No & Yes & Yes & Yes \\ \hline
IEEE C37.118.2 Compliance & Yes & Yes & Yes & No & Yes\\ \hline
Communication & Wired & Wired & Wired & Wired & Wireless (LTE cat-M) \\ \hline
\end{tabular}}
\label{table:Accuracy Comparison}
\end{table*}
LTE cat-M was originally introduced by 3GPP in Release 13 with a reduced bandwidth with respect to legacy LTE, coverage enhancement modes, support for half-duplex communications, and power reduction techniques, such as enhanced discontinuous reception. It was later improved with mobility enhancements (Rel. 14), improved spectral efficiency (Rel. 15), and ultra-low power wake-up radio operation to extend \ac{UE} battery life (Rel. 16).
The performance of LTE cat-M has been investigated by a large body of literature \cite{elsaadany17a, masek19,ratasuk17,hsieh18,medinaacosta19,dawaliby16}. 
In particular, Elsaadany et al. presented an overview on the features and challenges of cellular networks for IoT applications \cite{elsaadany17a}, focusing on the characteristics of the physical layer of LTE cat-M and exploring the pros and cons of this technology compared to available alternatives, such as LTE cat-0 and NB-IoT. \cite{masek19,ratasuk17,dawaliby16} used NS-3 to perform simulation at the network layer: \cite{masek19} analyzed a large-scale scenario with 500 end devices, however no indication of the type of traffic generated by each device was provided; \cite{ratasuk17} focused on the impact of network parameters, such as the number of repetitions, on the error rate, coverage, and battery life; \cite{dawaliby16} proposed a comparison of network performance with LTE cat-0 and observed a larger achieved throughput, with lower delay, frame losses and jitter with respect to LTE cat-0. \cite{medinaacosta19} evaluated the possibility of allocating fractions of resource blocks to \acp{UE} with small frames to transmit, as in the 5G massive machine-type communication scenario. An experimental study to analyze the coverage enhancement of LTE cat-M with respect to legacy LTE was proposed in \cite{hsieh18}. Despite the realistic setup considered in this study, also complemented with simulation results using Matlab, no information was provided on latency nor on the type of traffic generated by the \acp{UE}. 

The deployment of \ac{PMU} prototypes has already been investigated by a good deal of research and a comprehensive literature review can be found in \cite{schofield18}. In particular, \cite{Mohapatra15} employed 3 micro-controllers for phasor measurement and reporting. Another example is the work done in the OpenPMU project \cite{openpmu} to create a low cost open-source PMU. OpenPMU V1 uses National Instrument hardware and LabView software, but OpenPMU V2 employs complete open-source hardware and software. In the case of GridTrak PMU, presented in  \cite{gridtrak}, two micro-controllers are used, one for Phasor generation and another for PPS generation.  Differently from the existing literature, where wired technologies and protocols, such as Ethernet, were adopted, for this work we deployed a low-cost \micropmu with the ability to transmit data using LTE cat-M. A comparison of our work against some of the existing low-cost research-oriented \ac{PMU} deployments is shown in Table~\ref{table:Accuracy Comparison}.  

\section{System Description}\label{section:system}

\begin{figure}[h!]
\centering
\resizebox{\columnwidth}{!}{
\includegraphics[width=9cm]{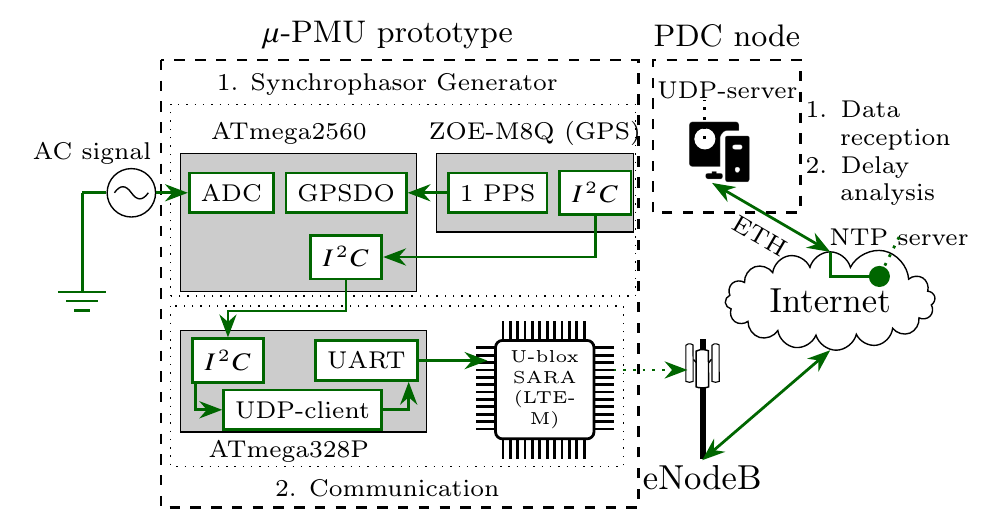}}
\caption{Schematic of the deployed testbed.}
\label{figure:system_schematic}
\end{figure}

The objectives of this study are to deploy a functional prototype of \micropmu with the ability to wirelessly transmit synchrophasor data to the \ac{PDC} and to study the suitability of LTE cat-M to transmit synchrophasor frames. In order to achieve these features, our \micropmu consists of two main modules, i.e., i) a synchrophasor generator module and ii) a communication module. The schematic of the system architecture is illustrated in Figure \ref{figure:system_schematic}, where the \micropmu prototype and the PDC node are enclosed in dashed boxes. The \micropmu box also contains two dotted boxes, which represent the two aforementioned modules. The PDC node is represented by a laptop connected to the Internet through an Ethernet cable. 
Our testbed also includes an AC signal generator (which can produce voltage signals with a 0-5V amplitude and a frequency in the range 1Hz-65534Hz) and a commercial LTE cat-M network with an eNodeB connected to the Internet\footnote{The core of the LTE cat-M network is not shown in Figure \ref{figure:system_schematic}, as it is not within the scope of this work.}.
In the remainder of this section, the system components are described in more detail.

\subsection{Synchrophasor generation module}\label{subsection:sync_generation}

In order to test the creation and transmission of realistic synchrophasor data, we need an AC signal to be monitored and an accurate time source. 
An Arduino ATmega2560 is employed to get the instantaneous voltage values from the AC signal generator and compute the synchrophasors. We can generally indicate with \synchrophasor ~the vector of synchrophasors measured at time $t_i$. In particular, \synchrophasor ~includes the voltage magnitude $v(t_i)$, phase $\phi(t_i)$, frequency $f(t_i)$ and \ac{ROCOF} $\rho(t_i)$. The system schematic for the synchrophasor generation module can be found in the upper portion of  Figure~\ref{figure:system_schematic}.
The baseline version of ATmega2560 mounts a 10-bit \ac{ADC}, which is used to sample the AC signals. The accuracy of the sampling timing is extremely important as any irregularity can undermine the successful computation of the power indicators in \synchrophasor. The sampling is controlled by an \ac{ISR}, which detects the rising edges of \ac{PWM} signals. \ac{PWM} signals are generated by a \ac{GPSDO}, implemented within the ATmega2560. The main function of the \ac{GPSDO} is to generate predefined, equidistant pulses synchronized with the detection of 1 pulse per second (PPS) from the ZOE-M8Q GPS module by means of an \ac{ISR}. Thus, a $1600$ Hz \ac{PWM} signal generated by the GPSDO will lead to a $1600$ Hz sampling by the ADC, which is enough to run a 32-point \ac{DFT} on a $50$ Hz signal. 
Different rates can be used to sample power signals at a different nominal frequency. For example, a frequency of $60$ Hz would require a sampling rate of $60 \cdot 32=1920$ Hz.

The GPS module, connected to the ATmega2560 over $I^2C$ interface, is used to maintain an accurate time synchronization. Fresh timing data is requested from the GPS at the start of every second, marked by the rising edge of the 1 PPS signal. The time between subsequent PPS peaks is calculated using internal timers of the micro-controller rendering an accuracy of 1 $\mu$s for time-stamping.  The primary reason for limiting timing requests to once per second only is due to the hardware limitations of the GPS module. The time-stamp of a given generated phasor \synchrophasor ~is denoted by $t_i$. 

As already indicated, a \micropmu must be able to measure the frequency $f(t_i)$ and calculate the ROCOF $\rho(t_i)$ of the AC signal. For that purpose, a sine to square wave converter circuit with a step down transformer is implemented, similarly to what was done in~\cite{Mohapatra15}. The AC signal is fed into the the circuit and the output is a square wave with voltage ranging from 0 V to 3 V. Another \ac{ISR} is used to keep track of the rising edges of the square wave to calculate the time difference between two consecutive rising edges, which represents the period $T$ of the signal. The frequency $f(t_i)$ is then calculated as the reciprocal of the period. The \ac{ROCOF} $\rho(t)$ was computed as $((f(t_i)-f_0)\cdot f_0)$, as shown in \cite{frigo19}.

A 32-point \ac{DFT} is used for estimating the magnitude $v(t)$ and phase $\phi(t)$  of the AC signal. A 32-element buffer is updated as soon as a new sample is generated by the ADC, which is controlled by the \ac{PWM} signal from the GPSDO. A counter keeps track of the number of samples and as soon as it reaches 32, \ac{DFT} is run. The power signal sampling and the subsequent \synchrophasor ~computation correspond to the first step of the \textit{Synchrophasor generation} phase, displayed on the timeline in Figure~\ref{figure:comm_workflow}. 
The second step in this phase is represented by the transmission (using the $I^2C$ interface) of synchrophasor \synchrophasor ~to the ATmega328P, which is responsible for the wireless transmission to the \ac{PDC} node.

\subsection{Communication module}\label{subsection:comm}

\begin{figure*}[h!]
\centering
\resizebox{2\columnwidth}{!}{
\includegraphics[scale=1]{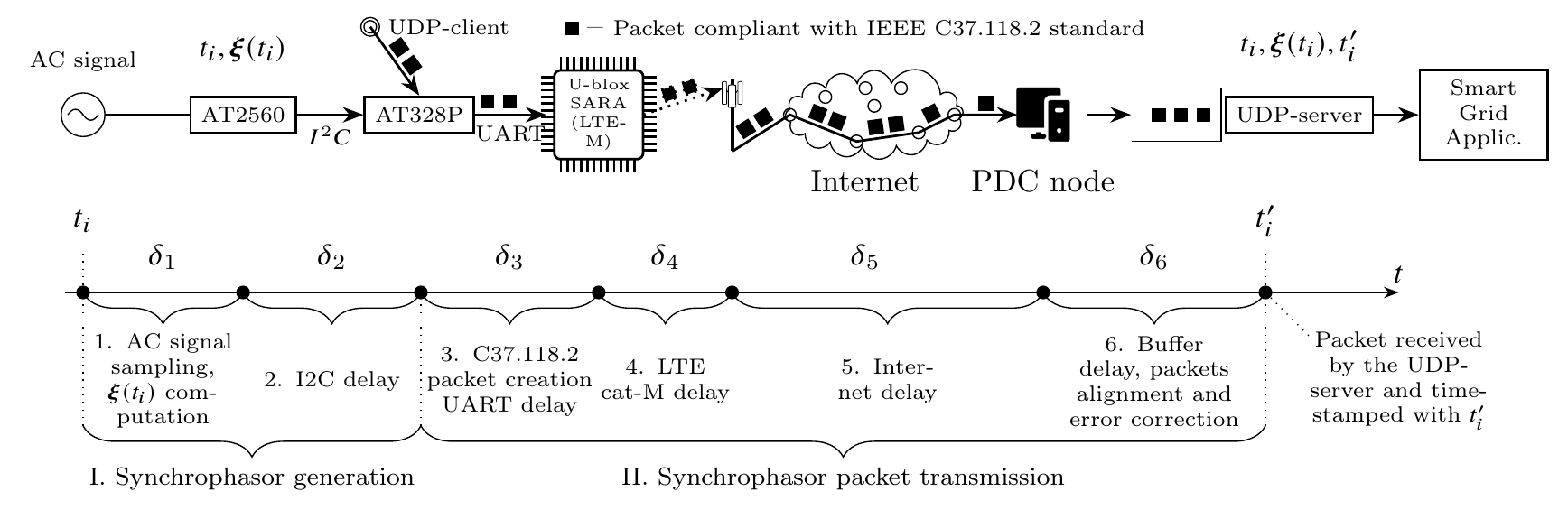}}
\caption{Workflow of the creation of synchrophasor frames and their transmission with the testbed.}
    \label{figure:comm_workflow}
\end{figure*}

Figure \ref{figure:comm_workflow} illustrates the workflow related to synchrophasor frames, represented here by black squares, from their creation in the ATmega328P device to their reception at the UDP-server hosted in the PDC node, which is represented by a laptop connected to the Internet. 
In particular, in the upper portion of the figure, we have a system schematic with the different components included in the testbed: they appear from left to right according to their sequential use in the workflow.
In the bottom part of Figure \ref{figure:comm_workflow}, it is possible to observe a timeline representing the two main phases in the workflow, i.e., I. synchrophasor  generation and II. synchrophasor frame transmission, and the six required steps.
The first two steps (included in Phase I) take place within the the Arduino ATmega2560 and were extensively described in the previous section. 
In order to perform the third and fourth steps (i.e., frame creation and frame transmission, respectively), we have used an ATmega328P, which is connected to the ATmega2560 over $I^2C$ interface. This interface between the synchrophasor generation module and the communication module is also visible in Figure \ref{figure:system_schematic}.

\begin{figure}[H]
\centering
\includegraphics[scale=1]{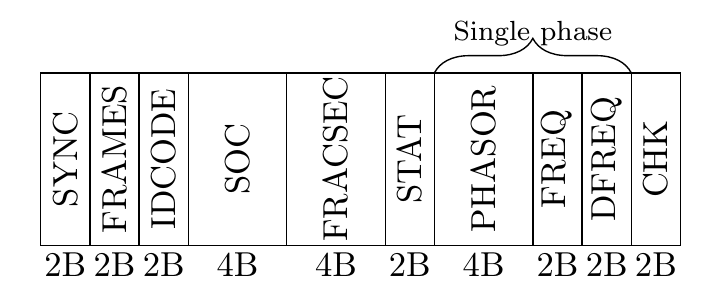}
\caption{Frame structure of a data frame in IEEE C37.118.2~\cite{IEEE_C37.118.2}.}
\label{fig:Packet Structure}
\end{figure}

Once a synchrophasor \synchrophasor ~is computed and timestamped with time $t_i$, the next step is to incorporate it in a frame, which corresponds to step 3 in Figure \ref{figure:comm_workflow}. In this work, we are following the IEEE C37.118.2 standard \cite{IEEE_C37.118.2}, whose main structure is shown in Figure \ref{fig:Packet Structure}: the frame size for a single phase system is 26 bytes and for a 3-phase system is 42 bytes. These values are related to a scenario with fixed 16-bit formats.
Among the frame fields, it is worth highlighting: (i) 
the Second of Century (SOC) 
and Fraction of Second (FRACSEC), which represent the time-stamp, (ii) 
PHASORS, which includes 
the estimated magnitude and phase of the monitored AC signal; (iii) FREQ, representing the operating frequency, and (iv) DFREQ, which represents the \ac{ROCOF} indicator.
The last field CHK is a checksum used for burst error detection according to the Cyclic Redundancy Check (CRC) technique  \cite{IEEE_C37.118.2}.
Further details on the IEEE C37.118.2 frame structure can be found in \cite{IEEE_C37.118.2,IEEE_27.244}. 

As shown in Figure \ref{figure:comm_workflow}, the transport protocol we used for the transmission of the synchrophasor frames is UDP, which assures low overhead and low latency communication as opposed to TCP, which is more suitable for reliable data transfers. A Python-based UDP client application was implemented in the ATmega328P to transmit synchrophasor frames and a Python-based UDP server was deployed at the PDC node to receive and collect these frames. Once synchrophasor frames are created, they are sent to the U-blox SARA module for thansmission over LTE cat-M. Frame creation and transmission over UART interface to the network module correspond to step 3 in Figure \ref{figure:comm_workflow}. The U-blox SARA module is then used to transmit frames to the LTE cat-M eNodeB (step 4). Frames are then transmitted over the Internet (step 5) until reaching the PDC node, where they are realigned\footnote{With UDP, frames can be received out-of-order.} and eventual errors are corrected using CRCs (step 6). Frames are finally received by the UDP server, which adds a time-stamp (i.e., $t_i'$). Received frames can be then used in a broad range of smart grid applications, as discussed in Section \ref{section:introduction}. The time stamp $t_i'$ is used in the end-to-end delay analysis provided in the next section.

\subsection{End-to-end delay and data analysis}\label{subsec:syncTx}

Once frames are received by the UDP server, they can be used to evaluate network performance and analyze the synchrophasor data.
One of the most important parameters to evaluate network performance (and assess the suitability of LTE cat-M for synchrophasor transmission) is the end-to-end delay, hereafter denoted as $\mathcal{D}(t_i)$, which is defined as the time elapsed from $t_i$, when \synchrophasor ~is generated, to $t_i'$, when \synchrophasor ~is received. Therefore, we have that $\mathcal{D}(t_i) = t_i'-t_i$. In Figure \ref{figure:comm_workflow}, it is shown that the reception time $t_i'$ is computed at the PDC, after the frame is correctly received by the UDP-server. However, in order to have meaningful and consistent values for $t_i$ and $t_i'$, the two devices measuring them, i.e., ATmega2560 and the laptop, respectively, need to be accurately synchronized. In order to achieve high precision synchronization, we used an NTP server, which is synchronized with the  the same GPS satellite used by our ZOE-M8Q GPS module. 

As shown in Figure \ref{figure:comm_workflow}, $\mathcal{D}(t_i)$ consists of the sum of the delays for each of the six steps, denoted by $\delta_k$ with $k=1,\dots,6$: In particular, $\delta_1$ represents the time needed to sample the power signal and compute \synchrophasor; $\delta_2$ represents the time needed to transmit \synchrophasor ~from the ATmega2560 to the ATmega328P over $I^2C$ interface; $\delta_3$ represents the time to create a frame and transmit it over the UART interface to the LTE cat-M card; $\delta_4$ and $\delta_5$ represent the network delay; $\delta_6$ represents the buffer delay, i.e., the time needed to sort frames that are received at the \ac{PDC} not in the correct order.

\subsection{Overview of LTE cat-M technology}\label{subsection:commu}


\begin{table*}
\centering
\small
\label{tab:table2}
\caption{Comparison of leading Communication Technologies}\cite{Dawaliby2016InDP}
\centering
\setlength{\tabcolsep}{3pt} 
\renewcommand{\arraystretch}{1.5} 
\resizebox{2\columnwidth}{!}{
\begin{tabular}{lllll}
\hline
& CAT-NB1 &  CAT-M1 & CAT-1 & LoRa \\ \hline
Peak Data Rate & 27.2/62.5 kb/s (DL/UL) &  375 kb/s (DL/UL) & 10/5 Mb/s (DL/UL) & 20 Kb/s (DL/UL)\\
Radio Spectrum & 180 KHz & 1.4 MHz & 20 MHz & 125 KHz\\
Latency & 1.5-10 s & 100-150 ms & 50-100 ms & High (Device Dependent)\\
Max. Coupling Loss & 164 dB & 155.7 dB & - & 165 dB\\
Tx Power & 46dBm/23dBm (DL/UL) & 43dBm/23dBm (DL/UL) & 23dBm (DL/UL) & 27dBm (DL/UL)\\
Noise Figure & 9dB/5dB (DL/UL) & 5dB/3dB (DL/UL) & 5dB (DL/UL) & 9dB (DL/UL)\\
Number of antennas & 1 & 1 & 2 & 1\\
VoLTE support & No & Yes & Yes & No\\
Power Consumption & LOW & LOW & HIGH & LOW\\
\hline
\end{tabular}
}
\label{table:Specification Comparison}
\end{table*}

The synchrophasors \synchrophasor, generated according to the mechanisms described in sections \ref{subsection:sync_generation} and \ref{subsection:comm}, need to be transmitted over a wireless network. In light of a steep increase of distributed generation elements to be integrated in the power distribution grid, a cheap and easy-to-install cellular IoT network is a good candidate to support this kind of high rate and small size communications.
The main alternatives in the cellular IoT technologies are LTE cat-M and NB-IoT\cite{chen17,li18}. Other communication technologies like LoRa and LTE CAT-1 should also be part of the discussion. The main reason behind not considering LTE CAT-1 for this application was the high power consumption and the main area where it excels, high data rate is not a requirement. On the other hand LoRa is highly power efficient but its low data rate and high latency makes it an in-viable option. A comparison between the standards is shown in Table~\ref{table:Specification Comparison}.
LTE cat-M was preferred over NB-IoT thanks to lower latency, higher peak data rate, and better coverage~\cite{CAT-M1_blog}. LTE cat-M is a low‑power wide‑area (LPWA) air interface that enables connection between IoT and M2M devices with medium data rate requirements. It can support up-link and down-link speeds of 375 kb/s in half duplex mode which is ideal for IoT applications requiring a low to medium data rate \cite{LTE-M}. 
Another benefit of LTE-M is "frequency hopping" among LTE sub-carriers making LTE-M more resistant to fading, interference as well as network congestion issues \cite{kosilo2020mobile}.
For this work, we use a commercial LTE cat-M network from a US-based carrier to transmit the synchrophasor data generated with the AT-mega2560. 



\section{Experimental Results}\label{section:results}

In this section, experimental results on the generation of synchrophasors \synchrophasor ~and on the performance of  the proposed communication solution are reported. 

\subsection{Synchrophasor generation}

\begin{figure}[h!]
	\subfloat[Magnitude plot]{
		\resizebox{\columnwidth}{!}{
		\includegraphics[]{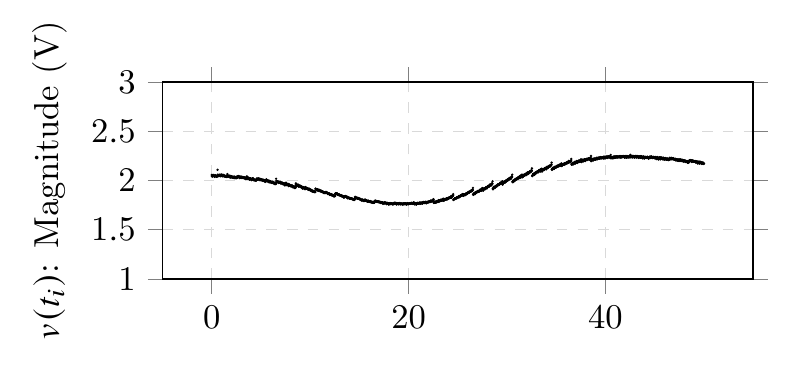}}
		\label{fig:Mag} 
	}\\\vspace{-.25cm}
	\subfloat[Phase plot]{
        \resizebox{\columnwidth}{!}{
		\includegraphics[]{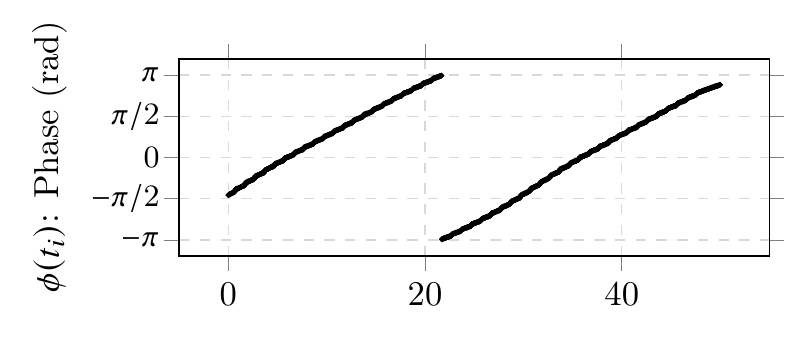}}
		\label{fig:Pha}
	}\\\vspace{-.5cm}
	\subfloat[Frequency plot]{
	    \resizebox{\columnwidth}{!}{
		\includegraphics[]{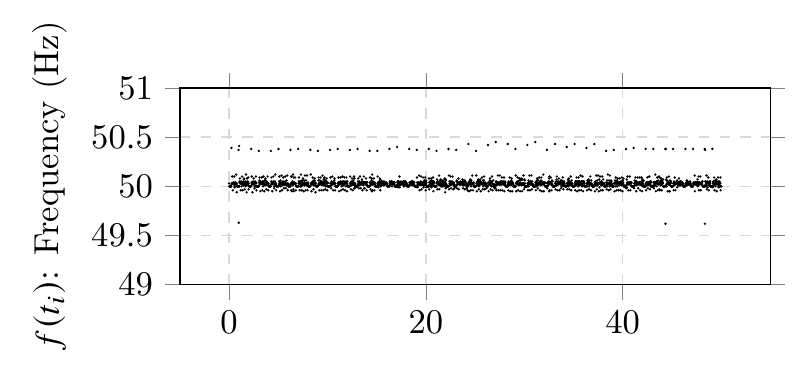}}
		\label{fig:Fre}
	}\\\vspace{-.5cm}
\captionsetup{captionskip=1pt}
	\subfloat[ROCOF plot]{
	   \resizebox{\columnwidth}{!}{
		\includegraphics[]{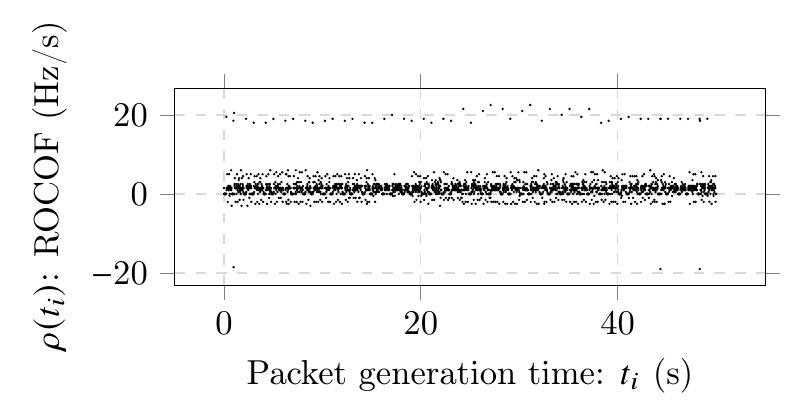}}
		\label{fig:ROCOF}
	}
	\vspace*{6mm}
	\caption{Synchrophasor characteristics with $f_0=50$ Hz, $v_0=2$ V, and a reporting rate of $50$ frames/s.}
	\label{fig:SC}
\end{figure}

Experimental results were obtained with the AC signal generator yielding a power signal with nominal values for magnitude and frequency of $v_0=2$ V and $f_0=50$ Hz, respectively. Synchrophasors \synchrophasor ~are computed and reported at a rate of $50$ frames/s and the obtained values are shown in Figure \ref{fig:SC}. This figure includes, from the top to the bottom, magnitude $v(t_i)$, phase $\phi(t_i)$, frequency $f(t_i)$ and \ac{ROCOF} $\rho(t_i)$, obtained in a $50$ s experiment. 

The frequency plot, depicted in Figure \subref*{fig:Fre}, shows that $f(t_i)$ fluctuates around the nominal value of $f_0=50$ Hz. These fluctuations can have several causes, such as imperfections in the signal generator device or in the sine to square wave converter circuit described in Section \ref{subsection:sync_generation}. 
In particular, the AC signal generator could operate at a frequency $f(t_i)\neq f_0$. As previously identified in \cite{Mohapatra15}, this fluctuation leads to incorrect DFT computation, whose sampling frequency is based on $f_0$, and causes (i) a fluctuation in the magnitude as a sinusoidal at a frequency $f(t_i)-f_0$, as shown in Figure \subref*{fig:Mag}, and (ii) a ``saw-tooth'' behaviour of the phase $\phi(t_i)$, confirmed by Figure \subref*{fig:Pha}, where $\phi(t_i)$ keeps on growing from $-\pi$ to $\pi$ when $f(t_i) > f_0$. Since $f(t_i)$ is predominantly  ($92\%$ of the observed values) greater than $f_0$ (see Figure \subref*{fig:Pha}), we can notice a similar trend to a sinusoidal in the magnitude plot in Figure \subref*{fig:Mag} and the expected saw-tooth behaviour in the phase plot in Figure \subref*{fig:Pha}. 
$\rho(t_i)$, computed as $(f(t_i)-f_0)\cdot f_0$ \cite{frigo19}, has a similar trend to $f(t_i)$ (Figure \subref*{fig:ROCOF}). This analysis shows the capability of the deployed \micropmu to estimate magnitude, phase, and frequency of a voltage waveform. In the remainder of this section, we provide a detailed analysis of the experimental LTE cat-M network performance results.

\vspace{-0.85cm}

\vspace{10pt}

\subsection{Network performance}

Network performance is extremely important to assess the feasibility of the proposed synchrophasor sensing and communication framework and to understand how to use the generated synchrophasor data.
In this section, we carry out a detailed analysis of relevant network \acp{KPI}, such as the delay $\mathcal{D}(t_i)$ and jitter. The delay $\mathcal{D}(t_i)$ has $6$ components $\delta_k$, as discussed in Section \ref{subsec:syncTx}. From our experimental results, $\delta_1$ and $\delta_3$ last less than 1 ms, whereas $\delta_2$ takes on average 1 to 2 ms. $\delta_6$ is equal to zero for almost all the synchrophasor frames in our experiments. Therefore, in what follows, we will focus on the LTE cat-M delay and on the Internet delay, which correspond to $\delta_4$ and $\delta_5$, respectively. 

As was illustrated in Figure \ref{figure:comm_workflow}, synchrophasors \synchrophasor ~are reported using an UDP socket: the client is located in the AT328P, and transmits the synchrophasors \synchrophasor ~at a fixed rate $\lambda$; the server is running at a laptop, connected to the Internet through a DSL router and an Ethernet cable. The server functionally represents the \ac{PDC} and is in charge of receiving the frames containing synchrophasor data, realign them, compensate for errors in the communication: these data are then available to be analyzed and used, as an example, for control and monitoring applications. Moreover, as previously mentioned, the \ac{PDC} must compute the delay $\mathcal{D}(t_i)$, which represents the time elapsed from the moment \synchrophasor ~is generated, i.e., $t_i$, and the time instant when \synchrophasor ~is received at the destination, i.e., $t_i'$. These delay values are stored at the laptop hosting the UDP server for post-processing and data analysis. The inter-transmission time between consecutive frames, hereafter denoted as $\Delta t$, can be easily computed as $\Delta t=t_i-t_{i-1}=1/\lambda$. 

In Table \ref{table:delay_stats}, statistics on the delay $\mathcal{D}_i$ obtained with our experimental setup are provided for different values of reporting rate $\lambda$, ranging from $1$ to $80$ frames/s. The most common values for $\lambda$ are $50$ and $60$ frames/s, however we decided to test the communication system with a broad range of reporting rates in order to gain a deeper insight on the suitability of LTE cat-M for a wide variety of smart grid applications. In Table \ref{table:delay_stats}, we included minima and maxima, standard deviations, first and third quartiles (i.e., $Q1$ and $Q3$), jitters, 95\%-confidence intervals of the true mean of the delay and frame loss.  
The average value of the delay, which can be found in the last column on the right, ranges from $162.95$ ms, obtained with $\lambda=80$ frames/s, to $173.50$ ms, obtained with $\lambda=60$ frames/s. It is worth pointing out that no frame loss or performance degradation is observed when the reporting rate grows. From this observation, it can be inferred that LTE cat-M is able to successfully and promptly deliver synchrophasor frames being transmitted at rates of at least up to $80$ frames per second. It is also worth highlighting that the standard deviation is very small, ranging from $17.05$ to $22.47$ ms, which also translates into narrow confidence intervals.
The main source of variability in the delay $\mathcal{D}_i$, as it will become evident in the rest of the analysis, comes from the access portion $\delta_4$.


\begin{table}[hbt!]
\caption{Delay Statistics Vs Reporting Rate.} 
\centering
\setlength{\tabcolsep}{3pt} 
\renewcommand{\arraystretch}{1.8} 
\resizebox{\columnwidth}{!}{
\begin{tabular}{c||cccccccc}
\multirow{2}{*}{\shortstack[l]{Report. rate\\$\lambda$ (frames/s)}} & \multicolumn{8}{c}{Delay $\mathcal{D}_i$ (ms)} \\
\cline{2-9}
& min & max & st.d. & Q1 & Q3 & jitter & 95\%-CI &  Fr. Loss\\
\hline\hline
1&145.37&201.17&20.19&147.22&187.29&40.02&$168.11\pm 2.29$&0\%\\
10&136.14&231.89&22.47&156.29&196.16&30.03&$168.78\pm 0.57$&0\%\\
50&138.47&514.56&18.34&159.93&184.75&19.61&$172.69\pm 0.29$ &0\%\\
60&136.69&246.34&17.96&157.33&188.12&18.36&$173.50\pm 0.26$ &0\%\\
80&132.97&226.44&17.05&148.86&177.91&14.14&$162.95\pm 0.22$ &0\%\\
\end{tabular}}
\label{table:delay_stats}
\end{table}


\begin{table}[hbt!]
\caption{Multi-PMU Delay Stats. ($\lambda=50Hz)$}  
\centering
\setlength{\tabcolsep}{3pt} 
\renewcommand{\arraystretch}{1.8} 
\resizebox{\columnwidth}{!}{
\begin{tabular}{c||cccccccc}
\multirow{2}{*}{\shortstack[l]{PMU\\num.}} & \multicolumn{8}{c}{Delay $\mathcal{D}_i$ (ms)} \\
\cline{2-9}
& min & max & st.d. & Q1 & Q3 & jitter & 95\%-CI &  Fr. Loss\\
\hline\hline
1 & 123.84 & 550 & 56.10 & 148.28 & 183.99 & 20.23 &$175.77 \pm 0.63 $ & 0\%\\
2 & 138.65 & 330 & 41.92 & 166.19 & 199.35 & 25.54 & $183.9 \pm 0.47 $ & 0\%\\
3 & 136.46 & 651 & 49.65 & 162.04 & 195.07 & 22.5 & $182.84 \pm 0.56 $ & 0\%\\
\end{tabular}}
\label{table:delay_stats_multi}
\end{table}

\begin{table}[ht!]
\caption{Delay Stats. Vs Frame Size ($\lambda=50Hz)$}
\setlength{\tabcolsep}{2.2pt}
\renewcommand{\arraystretch}{1.8} 
\resizebox{\columnwidth}{!}{
\begin{tabular}{c||ccccccc||c}
\multirow{2}{*}{\begin{tabular}[c]{@{}c@{}}Frame Size\\ (bytes)\end{tabular}} & \multicolumn{7}{c||}{Delay (ms)}                                    & \multirow{2}{*}{ Loss} \\ \cline{2-8}
                                                                              & min    & max     & st.d.  & Q1     & Q3     & jitter & 95\%-CI     &                              \\ \hline\hline
26 (1-Phase)                                                                   & 138.47 & 514.56  & 18.34  & 159.93 & 184.75 & 19.61  & $172.69\pm0.29$ & \multicolumn{1}{c}{0\%}   \\
42 (3-Phase)                                                                   & 160.64 & 293.08  & 14.61  & 182.20 & 204.73 & 16.83  & $193.99\pm0.23$ & \multicolumn{1}{c}{0.2\%}  \\
52 (2x)                                                                            & 164.11 & 259 & 17.29 & 158.9 & 183 & 31.67  & $170.88\pm0.27$ & \multicolumn{1}{c}{0.6\%}  \\
78  (3x)                                                                           & 168.14 & 367.62  & 20.37  & 195.96 & 226.41 & 30.14  & $211.38\pm0.32$ & \multicolumn{1}{c}{3.3\%}  
\end{tabular}}
\label{table:delay_stats_size}
\end{table}

\begin{figure}[H]
\centering
\resizebox{\columnwidth}{!}{
\includegraphics[]{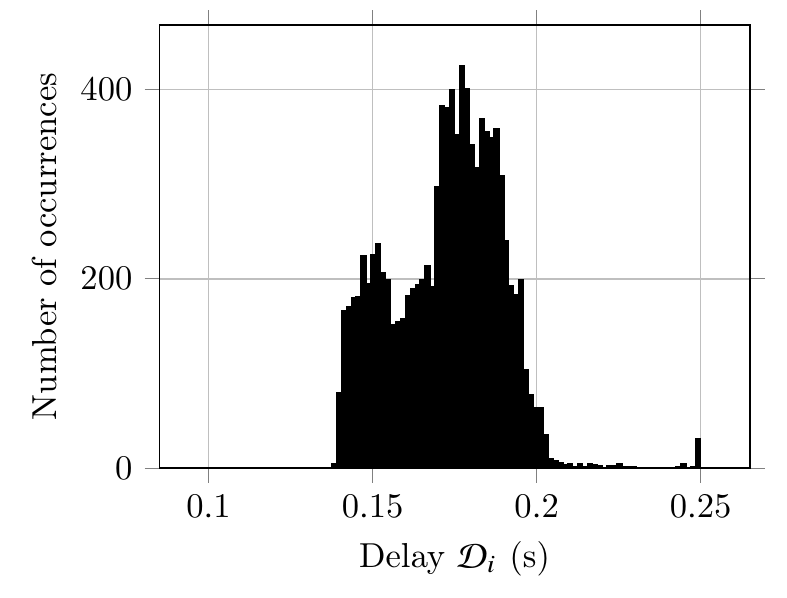}
}
\caption{Histogram of the delays obtained with $\lambda=50$ frames/s.}
\label{fig:Packet Delay Density Histogram}
\end{figure}

The distribution of the delay obtained with $\lambda=50$ frames/s is shown in Figure \ref{fig:Packet Delay Density Histogram}. Delays are distributed in the interval between 140 and 250 ms, even though approximately $98$\% of frames is received within $200$ ms. The maximum value of $514$ ms, reported in Table \ref{table:delay_stats}, is not shown in the distribution graph as it is a single value obtained in the first frame transmission (due to to the time needed to establish the connection), which makes it irrelevant to the network performance analysis. These results further corroborates the finding that LTE cat-M is a suitable communication solution to transmit synchrophasor data.

A multi-PMU setup and framesize variation were also considered as a part of the analysis. Table \ref{table:delay_stats_multi} depicts the frame statistics of a 3 PMU setup which transmits simultaneously at $50 hz$. The 95\%-confidence interval of the delay as well as the frame loss is consistent with the results obtained before. Table \ref{table:delay_stats_size} shows the impact of frame size variation on the frame statistics, where single phase (26 bytes), 2x single phase (52 bytes), 3x single phase (78 bytes) and 3-Phase (42 bytes) frames were considered. We observed that an increase in the frame size corresponds to increased frame losses. However, the delay is not majorly impacted and, as indicated by the 95\% CI, stays consistently within the 140-250ms range. 

\begin{figure}[h!]
\centering
\resizebox{\columnwidth}{!}{
\includegraphics[]{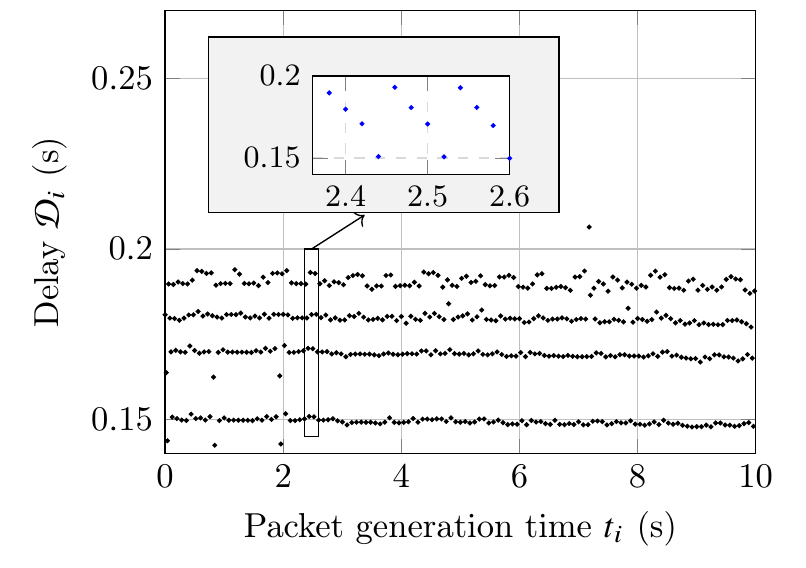}
}
\caption{Delay $\mathcal{D}_i$ with respect to the frame generation time $t_i$ in an experiment with $\lambda=50$ frames/s.}
\label{fig:delay_over_time}
\end{figure}

Further insight into the network performance can be gained analyzing Figure \ref{fig:delay_over_time}, which displays how the delay $D_i$ evolves with respect to the transmission time $t_i$ in a scenario with $\lambda=50$ Hz. 
In the $10$ s observation window reported in this figure, it is possible to notice that the delays are not uniformly distributed in the range between $150$ ms and 200 ms but are concentrated in the vicinity of $150$, $170$, $180$, and $190$ ms. A zoom with the delay of 12 consecutive synchrophasor frames is depicted in a box on top of Figure \ref{fig:delay_over_time}. A pattern is identified for each group of four consecutive synchrophasor frames, whose delay decreases from the first to the last. In particular, in each group, the first frame is received with a delay of around $190$ ms, the second with a delay of around $180$ ms, the third with a delay of around $170$ ms, and the fourth and last with a delay of around $150$ ms. This behaviour is observed throughout the whole experiment with $\lambda=50$ synchrophasors per second.

In order to explain the reason for such a pattern in the observed delays, it is important to consider the mechanism used to allocate radio resources in LTE cat-M networks.
In particular, before each frame transmission, the UE must send a \textit{scheduling request} control message to the eNodeB, and must wait for a scheduling grant, which indicates the time and frequency resources to be used for frame transmission. However, scheduling grants are transmitted, along with other data, in the system information (SI) with a periodicity that depends on the network parameter \textit{si-WindowLength-BR-r13}, which could range from $20$ ms to $200$ ms \cite{3gpp.36.331}. In the commercial network used to obtain our results, this parameter is equal to $80$ ms. In the scenario at hand, we have a reporting rate $\lambda=50$ frames/s, which leads to an inter-departure time $\Delta t=1/\lambda=20$ ms. Therefore, 4 frames are expected to be transmitted between two consecutive SI frames. In each 4-frame group, the delay decreases from the first to the last due to decreasing waiting time until the following SI transmission.

\captionsetup{captionskip=4pt}
\begin{figure}[h!]
	\centering
	\subfloat[$\lambda=50$ frames/s]{
		\resizebox{\columnwidth}{!}{
		\includegraphics[]{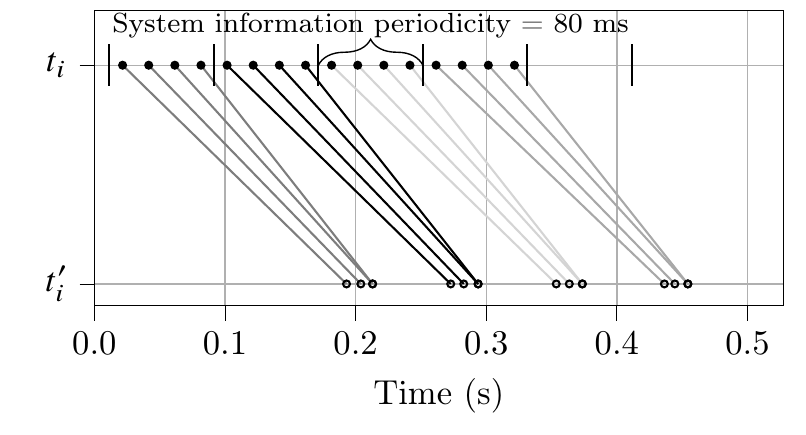}}
        \label{fig:PV_50}
	}\\\vspace{-0.25cm}
	\subfloat[$\lambda=12.5$ frames/s]{
        \resizebox{\columnwidth}{!}{
		\includegraphics[]{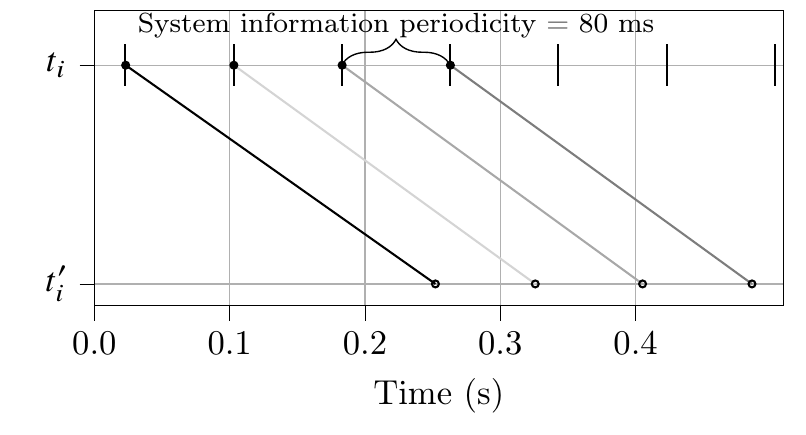}}
		\label{fig:PV_12}
	}
	\caption{Timeline with generation time $t_i$ and the reception time $t_i'$ for synchrophasor frames with (a) $\lambda=50$ frames/s and (b) $\lambda=12.5$ frames/s.}
\label{fig:PV}
\end{figure}

To better visualize this effect, in Figure \ref{fig:PV}, we have reported the generation time $t_i$ and the reception time $t_i'$ for a number of consecutive synchrophasor frames. It is worth highlighting that $t_i$ is very close to the transmission time of synchrophasor frame, as $\delta_1$, $\delta_2$ and $\delta_3$ are negligible, as discussed in Section \ref{subsec:syncTx}. Note also that each synchrophasor frame is represented by a segment with the starting point in the $t_i$-timeline at the top and the ending point in the $t_i'$-timeline at the bottom. For the experiment with $\lambda=50$ frames/s, groups of 4 frames are observed in Figure \subref*{fig:PV_50}, and the segments representing frames in the same group are drawn with the same color in the gray scale. It is worth highlighting the differences in time-distributions of frame transmissions (represented by filled circles on the top timeline) and of the frame receptions (represented by empty circles on the bottom timeline): frame transmission times are equally spaced by $80$ ms, whereas the reception times of frames in the same group are just a few ms apart. It is also possible to note that the delay of frames in each group decreases from the first, which has the longest waiting time to the following SI transmission, to the last one, which is the one with the lowest waiting time to the following SI transmission. In Figure \subref*{fig:PV_12}, a similar plot was generated for a scenario with $\lambda=12.5$ frames/s, where a considerably more regular distribution of frame arrival times can be observed. This is due to the fact that frames are transmitted at an inter-departure time of $\Delta t=1/12.5=80$ ms, which is equal to the SI periodicity parameter: this means that frames wait, on average, the same time before being transmitted over LTE cat-M.

\section{Conclusions}\label{section:conclusions}

In this paper, we presented research on the experimental deployment of a $\mu$-PMU device and an analysis of the feasibility of using the LTE cat-M network to support high-rate data collection from these devices. A functional prototype of a \micropmu was deployed using simple hardware, such as Arduino micro-controllers and GPS antennas. The deployed \micropmu was equipped with a wireless card to transmit over an LTE cat-M commercial network. 

A wide variety of reporting rates for synchrophasor data (i.e., 1 to 80 frames/s) was considered in order to represent the traffic generated by a broad range of potential smart grid use cases, such as protection and control applications, state estimation, and outage management. Synchrophasor data were generated in compliance with the widely adopted IEEE C37.118.2 standard.

A detailed LTE cat-M network performance analysis was carried out, with average delays below $200$ ms in all the considered scenarios, making LTE cat-M particularly suitable to support this type of traffic. Finally, the effects of network parameters, such as the system information periodicity, on the network performance was studied. It was shown that this parameter considerably affects the time distribution of the frame arrivals, which is strikingly different from the periodic distribution of the frame generations.

\section*{Acknowledgment}

The authors would like to thank Adarsh Govindarajan for his help with the deployment of the sine to square wave converter circuit.

\bibliographystyle{IEEEtran}
\bibliography{main}

\end{document}